\documentclass[12pt]{article}
\usepackage{amsfonts}

\begin{document}
\begin{titlepage}

\begin{center}


{\Large \bf Spin-1 gravitational waves}\\
\vspace{1cm}
{\bf F. Canfora}$^a$, {\bf G. Vilasi}$^a$, {\bf P. Vitale}$^b$\\[6mm]
 {\it $^a$Dipartimento di Fisica ''E.R.Caianiello''},
{\it Universit\`{a} di Salerno}\\ {\it and INFN, Gruppo Collegato
di Salerno,
Salerno, Italy.}\\
{\it $^b$Dipartimento di Scienze Fisiche, Universit\`a di Napoli  Federico II} \\
{\it and INFN Sezione di Napoli, Napoli, Italy.}\\
{\tt canfora@sa.infn.it, vilasi@sa.infn.it, vitale@na.infn.it}
\end{center}


\begin{abstract}
Gravitational fields invariant for a $2$-dimensional Lie algebra
of Killing fields $\left[ X,Y\right] =Y$, with $Y$ of
\textit{light type,} are analyzed. The conditions for them to
represent gravitational waves are verified and the definition of
energy and polarization is addressed; realistic generating sources
are described.
\end{abstract}

\end{titlepage}


\section*{Introduction}

Gravitational waves, that is a propagating warpage of space time
generated from compact concentrations of energy, like neutron
stars and black holes, have not yet been detected directly,
although their indirect influence has been seen and measured with
great accuracy. Presently there are, worldwide, many efforts to
detect gravitational radiation, not only because a direct
confirmation of their existence is interesting \textit{per se} but
also because new insights on the nature of gravity and of the
Universe itself could be gained. For these reasons exact solutions
of the Einstein field equations deserve special attention when
they are of propagative nature. The need of taking into full
account the nonlinearity of Einstein's equations when studying the
generation of gravitational waves from strong sources is generally
recognized \cite{Th95}. Moreover, despite the great distance of
the sources from Earth (where almost all the experimental devices,
laser interferometers and resonant antennas, are located) there
are situations where the non linear effects cannot be neglected.
This is the case when the source is a binary coalescence: indeed
it has been shown \cite{Ch91} that a secondary wave, called the
\textit{Christodoulou memory} is generated via the non linearity
of Einstein's field equations. The memory seems to be too weak to
be detected from the present generation of interferometers \cite
{Th95} (even if its frequency is in the optimal band for the
LIGO/VIRGO interferometers) but of the same order as the linear
effects related to the same source, thus stressing the relevance
of the nonlinearity of the Einstein's equations also (soon) from
an experimental point of view.

On the theoretical side, starting from the seventy's new powerful
mathematical methods have been developed to deal with nonlinear evolution
equations. For instance, a suitable generalization of the \textit{\ Inverse
Scattering Transform }allows to integrate \cite{BZ78} Einstein field
equations for a metric of the form
\[
g=f\left( z,t\right) \left( dt^{2}-dz^{2}\right) +h_{11}\left( z,t\right)
dx^{2}+h_{22}\left( z,t\right) dy^{2}+2h_{12}\left( z,t\right) dxdy.
\]
Indeed, the corresponding vacuum Einstein field equations reduce essentially%
\footnote{%
The function $f$ can be obtained by quadratures in terms of the matrix $%
\mathbf{\ \ H}$.} to
\[
\left( \alpha \mathbf{H}^{-1}\mathbf{H}_{\xi }\right) _{\eta }+\left( \alpha
\mathbf{\ H}^{-1}\mathbf{H}_{\eta }\right) _{\xi }=0,
\]
where $\mathbf{H}\equiv $ $\left\| h_{ab}\right\| ,$ $~\xi =\left(
t+z\right) /\sqrt{2},\quad \eta =\left( t-z\right) /\sqrt{2},\quad \alpha =%
\sqrt{\left| \det \mathbf{H}\right| }$. This is a system of
non-linear differential equations whose form is typical for
two-dimensional integrable systems. Its solution through the
\textit{\ Inverse
Scattering Transform}, yields \textit{%
\ gravitational solitary waves solutions}.

A geometric inspection of the metric above shows that it is invariant under
translations along the $x,y$-axes, \textit{i.e. }it admits two Killing
fields, $\partial _{x}$ and $\partial _{y}$, closing on an Abelian\footnote{%
The study of metrics invariant for a Abelian $2$-dimensional
Killing Lie algebra goes back to Einstein and Rosen
\cite{ER37,Ro54}, Kompaneyets \cite {Ko58}; recent analysis can be
found in \cite{Ve93,NN01}} two-dimensional Lie algebra
$\mathcal{A}_{2}$. Moreover, the distribution $\mathcal{D}$,
generated by ${\partial _{x}}$ and ${\partial _{y}}$, is
$2$-dimensional and the distribution $\mathcal{D}^{\mathcal{\perp
}}$ orthogonal to $\mathcal{D}$ is integrable and transversal to
$\mathcal{D} $.

Thus, it has been natural to consider \cite{SVV00} the general problem of
characterizing all gravitational fields $g$ admitting a Lie algebra $%
\mathcal{G}$ of Killing fields such that:

I. the distribution $\mathcal{D}$, generated by the vector fields of $%
\mathcal{\ G}$, is two-dimensional.

II. the distribution $\mathcal{D}^{\mathcal{\perp }}$ orthogonal to $%
\mathcal{D}$ is integrable and transversal to $\mathcal{D}$.

The aim of this article is to study, among those solutions, the
ones which show a propagative nature. A preliminary account of our
results is given in \cite{CVV02}.

The article is organized as follows. In section \textbf{1}
gravitational fields invariant for a two dimensional Lie algebra
are characterized by reducing the Einstein equations to the so
called $\mu $-deformed Laplace equation. Harmonic coordinates are
also introduced. Section \textbf{2} is devoted to the analysis of
the wave-like character of the solutions through the Zel'manov and
the Pirani criteria. In section \textbf{3} realistic sources are
described. In section \textbf{4} Landau-Lifshitz's and Bel's
energy-momentum pseudo-tensors are introduced and a comparison
with the linearised theory is performed. Eventually, section
\textbf{5} is devoted to the analysis of the polarization of the
waves.

In the following, $\mathcal{K}il\left( g\right) $ will denote the Lie
algebra of all Killing fields of a metric $g$ while \textit{Killing algebra}
will denote a sub-algebra of $\mathcal{K}il\left( g\right) $.

Moreover, an integral (two-dimensional) submanifold of
$\mathcal{D}$ will be called a \textit{Killing leaf}, and an
integral (two-dimensional) submanifold of $\mathcal{D}^{\bot }$
\textit{orthogonal leaf.}

\section{Invariant vacuum gravitational fields}

Let $g$ be a metric on the space-time $M$ and $\mathcal{G}_{2}\,\ $one of
its Killing algebras whose generators $X,Y$ satisfy the commutation relation
\begin{equation}
\left[ X,Y\right] =\sigma Y,\,\,\,\,\sigma =0,1.  \label{kil}
\end{equation}
The Frobenius distribution $\mathcal{D}$ is two-dimensional and a coordinate
system $(x^{\mu })$, $\mu =1,2,3,4$ exists such that
\[
X=\frac{\partial }{\partial x^{3}},\,\,\ \,Y=\exp \left( \sigma x^{3}\right)
\frac{\partial }{\partial x^{4}}.
\]
Such a coordinate system is called \cite{SVV00} \textit{semiadapted} (to the
Killing fields).

Condition II of the previous section allows to construct special
semi-adapted charts such that the fields $e_{1}=\partial /\partial x^{1}$, $%
e_{2}=\partial /\partial x^{2}$, belong to $\mathcal{D}^{\bot }$. In such a
chart, called from now on \textit{adapted}, the most general $\mathcal{G}%
_{2} $-invariant metric has the form \cite{SVV00}
\begin{eqnarray*}
g &=&g_{ij}dx^{i}dx^{j}+\left( \sigma ^{2}\lambda \left( x^{4}\right)
^{2}-2\sigma \mu x^{4}+\nu \right) dx^{3}dx^{3}+ \\
&&2\left( \mu -\sigma \lambda x^{4}\right) dx^{3}dx^{4}+\lambda
dx^{4}dx^{4},\quad i,j=1,2
\end{eqnarray*}
where\ $g_{ij}$, $\lambda $, $\mu $, $\nu $ are arbitrary functions of $%
\left( x^{1},x^{2}\right) $.

If the Killing field $Y$ is not of \textit{\ light type}, i.e.
$g(Y,Y)\neq 0$ , condition II follows automatically from I. The
local structure of this class of Einstein metrics has been
explicitly described; \
it turns out \cite{SVV00} that a third Killing field $Z$ exists such that $%
\left( X,Y,Z\right) $ span the Killing algebra $so\left( 2,1\right) $ and
all invariant metrics are static and locally diffeomorphic to the \textit{%
pseudo-Schwarzschild metric. }This metric was also found in the context of
warped solutions \cite{KKK99,VV01}.

In the following the Killing field $Y$ will be assumed to be of \textit{\
light type}. In this case the general solution of vacuum Einstein equations,
in the adapted coordinates $\left( x^{1},x^{2},x^{3},x^{4}\right) $, is
given by
\begin{equation}
g=2f(dx^{1}dx^{1}+dx^{2}dx^{2})+\mu \lbrack (w\left( x^{1},x^{2}\right)
-2\sigma x^{4})dx^{3}dx^{3}+2dx^{3}dx^{4}],  \label{gm}
\end{equation}
where $\mu =D\Phi +B$; $D,B\in \mathcal{R}$, $\Phi $ is a non constant
harmonic function of $x^{1}$ and $x^{2}$, $f=\left( \nabla \Phi \right) ^{2}%
\sqrt{\left| \mu \right| }/\mu $, and $w\left( x^{1},x^{2}\right) $ is a
solution of the $\mu $-\textit{deformed Laplace equation:}
\begin{equation}
\Delta w+\left( \partial _{x^{1}}\ln \left| \mu \right| \right) \partial
_{x^{1}}w+\left( \partial _{x^{2}}\ln \left| \mu \right| \right) \partial
_{x^{2}}w=0.  \label{dle}
\end{equation}
The solutions of the $\mu $-deformed Laplace equation will also be
called \textit{\ }$\mu $-\textit{harmonic functions.}

In the particular case $\sigma =1$, $f=1/2$ and $\mu =1$, the above metrics
are locally diffeomorphic to a subclass of the vacuum Peres solutions \cite
{Pe59} corresponding to a special choice of the harmonic function
parameterising that metrics.

This is easily seen by introducing, for $\widetilde{u}\neq 0$, new
coordinates $\left( x,y,\widetilde{u},\widetilde{v}\right) $\
defined by $ x^{1}=x,\,\ \ \,x^{2}=y,\,\ \ \ x^{3}=\ln \left|
\widetilde{u}\right| ,\,\ \ \ x^{4}=\widetilde{u}\widetilde{v}. $

Of course, metrics (\ref{gm}) can also be obtained by imposing an additional
isometry to metrics admitting a Killing vector field of light type and,
indeed, they are an extension of the ones reported in \cite{KSHM80}.

When $w$ is constant the above family of Einstein metrics admits the \textit{%
\ time-like} Killing vector field
\[
T=\exp \left( -x^{3}\right) \frac{\partial }{\partial x^{3}}+\frac{1}{2}%
\left[ \left( w-2\sigma x^{4}\right) \exp \left( -x^{3}\right) +\exp \left(
x^{3}\right) \right] \frac{\partial }{\partial x^{4}}
\]
which is \textit{hypesurfaces orthogonal}. This means that these Einstein
metrics are just \textit{static} gravitational fields.

As it will be argued in the next sections, when $w$ is not constant the
above family of Einstein metrics may represent propagative gravitational
fields.

Since the distribution $\mathcal{D}^{\bot }$\ is assumed to be transversal
to $\mathcal{D}$, the restriction of $g$\ to any Killing leaf, say $\mathcal{%
S}$, is non-degenerate. So, $\left( \mathcal{S},\left. g\right| _{S}\right) $
\ is a homogeneous two-dimensional Riemannian manifold. In particular, the
Gauss curvature $K\left( \mathcal{S}\right) $ \ of the Killing leaves is
constant. An explicit computation shows that $K\left( \mathcal{S}\right) $
vanishes.

Thus, the space-time $\mathcal{M}$ has a fiber bundle structure
\[
\pi :\mathcal{M}\longrightarrow \mathcal{W},
\]
whose basis $\mathcal{W}$\ is diffeomorphic to the orthogonal
leaves and whose fibers are the Killing leaves and as such are
flat two-dimensional Riemann manifolds.

Despite the non-linear nature of general relativity, gravitational
fields\ (\ref{gm}) obey to two superposition laws. Indeed, with
\textit{two}
harmonic functions $\Phi _{1}$ and $\Phi _{2}$ we can associate \textit{three%
} gravitational fields (in facts a whole two-parameters family), that is, $%
g_{\Phi _{1}}$, $g_{\Phi _{2}}$ and $g_{a\Phi _{1}+b\Phi _{2}}$;
the last
one, which is associated with the linear combination of $\Phi _{1}$ and $%
\Phi _{2}$, may be regarded as the superposition of the two
associated solutions $g_{\Phi _{1}}$ and $g_{\Phi _{2}}$.

The second superposition law follows from the linearity of the
$\mu $-deformed Laplace equation, so that with two $\mu
$-\textit{harmonic} functions $w_{1}$ and $w_{2}$ we can associate
\textit{three} gravitational fields $g_{w_{1}}$, $g_{w_{2}}$ and
their {\it sum} $g_{w_{1}+w_{2}} \equiv \left(
g_{2w_{1}}+g_{2w_{2}}\right) /2$

\subsection{\protect\smallskip The harmonic coordinates.}

The coordinates $(x^{3},x^{4})$ on the Killing leaves $\mathcal{S}$ have a
clear geometric meaning but are of difficult physical interpretation.
Fortunately, being the Killing leaves flat manifolds, it is possible to
introduce coordinates $(\widetilde{z},\widetilde{t})$ diagonalizing the
metric
\[
\left. g\right| _{\mathcal{S}}=\widetilde{\mu }[(\widetilde{w}-2\sigma
x^{4})dx^{3}dx^{3}+2dx^{3}dx^{4}],
\]
where $\widetilde{\mu }$ and $\widetilde{w}$, being the restriction of the
functions $\mu $ and $w$ to the Killing leaves, are constant.

The coordinate system $\left( x,y,z,t\right) $, where for $z>t$
\[
\left\{
\begin{array}{l}
x=w_{1}\left( x^{1},x^{2}\right) \\
y=w_{2}\left( x^{1},x^{2}\right) \\
z=\frac{1}{2}\left[ \left( 2x^{4}-w\left( x^{1},x^{2}\right) \right) \exp
\left( x^{3}\right) +\exp \left( -x^{3}\right) \right] \\
t=\frac{1}{2}\left[ \left( 2x^{4}-w\left( x^{1},x^{2}\right) \right) \exp
\left( x^{3}\right) -\exp \left( -x^{3}\right) \right] ,
\end{array}
\right.
\]
is \textit{harmonic}, $w_{1}\left( x^{1},x^{2}\right) $ and $w_{2}\left(
x^{1},x^{2}\right) $ denoting any two independent $\mu $%
-harmonic functions. The generic Killing leaf $\mathcal{S}$ is mapped onto
the half-plane $z>t$, the line $z=t$ representing the points with $%
x^{3}=+\infty $.

In these coordinates, metrics (\ref{gm}) take the form
\begin{eqnarray}
g &=&2\frac{\sqrt{\mu }\left( \nabla \Phi \right) ^{2}J^{-2}}{\mu }[(\nabla
y)^{2}dx^{2}+(\nabla x)^{2}dy^{2}-2\nabla x{\ensuremath{\cdot}}\nabla ydxdy]
\label{ggen} \\
&&+\mu \lbrack dz^{2}-dt^{2}+d\left( w\right) d\left( \ln \left|
z-t\right| ]\right)],  \nonumber
\end{eqnarray}
where $J=\partial _{x^{1}}w_{1}\partial _{x^{2}}w_{2}-\partial
_{x^{2}}w_{1}\partial _{x^{1}}w_{2}$ is the Jacobian determinant of the map $%
\left( x^{1},x^{2}\right) \rightarrow \left( x,y\right) $.

In the case $\mu =$ \textit{const}, the $\mu $-deformed Laplace equation
reduces to the Laplace equation and $w_{1}$, $w_{2}$ reduce to be just
harmonic functions. Thus, it is possible to choose $x=x^{1},y=x^{2}$ so that
in the harmonic coordinates $\left( x,y,z,t\right) $, and for $\mu =1$, the
above Einstein metrics take the particularly simple form
\begin{equation}
g=2f(dx^{2}+dy^{2})+dz^{2}-dt^{2}+d\left( w\right) d\left( \ln
\left| z-t\right| ]\right) .  \label{gw}
\end{equation}
This coordinates system explicitly shows that, when $w$ is constant, the
Einstein metrics given by Eq. (\ref{gw}) are static and, under the further
assumption $\Phi =x\sqrt{2}$, they reduce to the Minkowski one. Moreover,
when $w$ is not constant, gravitational fields (\ref{gw}) look like a
\textit{disturbance} moving, along the$\ z$ direction on the Killing leaves,
at light velocity. However, the last observation is neither rigorous nor
covariant. Since the propagation direction is the most important ingredient
in the study of the polarization, in the following sections a detailed
analysis will be devoted to this question.

In the following we will assume that $w$ is not constant.

\section{Zelmanov's and Pirani's criteria}

To check the wave character of gravitational fields (\ref{gm}), a
Zakharov generalization of the Zel'manov criterion \cite{Za72}
will be applied which states that a vacuum solution of the
Einstein equations is a gravitational wave if the components\
$R_{\mu \nu \lambda \sigma }$ of the corresponding Riemann tensor
field $R$, satisfy a hyperbolic equation of the form
\begin{equation}
g^{\alpha \beta }\nabla _{\alpha }\nabla _{\beta }R_{\mu \nu \lambda \sigma
}=N_{\mu \nu \lambda \sigma }  \label{zel}
\end{equation}
where $\ \nabla _{\beta }$ denotes the Levi-Civita covariant derivative of
the metric and $N_{\mu \nu \lambda \sigma }$ denote the components of a
tensor field $N$ depending at most on first derivatives of the Riemann
tensor itself.

For \textit{symmetric manifolds} Eq. (\ref{zel}) with $N=0$ is an identity
because the Riemann tensor is covariantly constant, but it may become an
identity also in the case of \textit{Einstein manifolds} ($R_{\alpha \beta
}=\kappa g_{\alpha \beta }$), for special choices of the tensor field $N$.
Hence, to exclude a priori these situations, the original Zel'manov
criterion is formulated in the more restrictive assumptions:

\begin{itemize}
\item  $R_{\alpha \beta \gamma \delta }$ not covariantly constant\footnote{%
That is the manifold is not symmetric.} ;

\item  $g^{\alpha \beta }\nabla _{\alpha }\nabla _{\beta }R_{\mu \nu \lambda
\sigma }=0$ .
\end{itemize}

The metrics in Eq. (\ref{gm}) certainly do not define symmetric or Einstein
manifolds, as can be checked from the components of the Ricci tensor given
below. Hence, the first hypothesis is certainly satisfied while the second
one, \textit{i.e. }$N=0$, which ensures the applicability of the criterion
to Einstein manifolds too, is not needed.

Concerning the physical meaning of this criterion, it can be shown
\cite {Za72} that the characteristic hypersurface of the system of
equations (\ref {zel}) is identical with the characteristic
hypersurface of the Einstein and Maxwell equations in curved
space-time. Consequently, Eqs. (\ref{zel}) describe the
propagation of the discontinuities of the second derivatives of
the Riemann tensor. This links the Zel'manov criterion to the
intuitive concept of \textit{local wave of curvature}. The
criterion is independent on the explicit form of $N_{\mu \nu
\lambda \sigma }$; in fact, the characteristic hypersurface of a
system of equations is determined only by
the highest derivative term. Then we will not fix an explicit form of $%
N_{\mu \nu \lambda \sigma }$ but just require that $N_{\mu \nu \lambda
\sigma }$ be a tensor containing at most first derivatives of the Riemann
tensor. This clearly corresponds to a covariant criterion \cite{Za72}. Then
a sufficient condition is
\begin{equation}
g^{\alpha \beta }\partial _{\alpha }\partial _{\beta }R_{\mu \nu
\lambda \sigma }=0  \label{gzel},
\end{equation}
where $\partial _{\beta }$ are the usual partial derivatives. In fact, if
this is the case then $N_{\mu \nu \lambda \sigma }$ is a tensor containing
at most first derivatives of the Riemann tensor.

To start with, let us verify that gravitational fields (\ref{gw})
do satisfy Eqs. (\ref{gzel}).

In the harmonic coordinates system the only nonvanishing components of the
Riemann and Ricci tensor fields corresponding to metrics (\ref{gw}) are
proportional to one of the following
\begin{eqnarray}
R_{txzx} &=&\frac{\left( 2fw,_{xx}\ +f,_{y}w,_{y}-f,_{x}w,_{x}\right) }{%
4f\left( z-t\right) ^{2}}  \nonumber \\
R_{txzy} &=&\frac{\left( 2fw,_{xy}-f,_{y}w,_{x}-f,_{x}w,_{y}\right) }{%
4f\left( z-t\right) ^{2}}  \nonumber \\
R_{tyzy} &=&\frac{\left( 2fw,_{yy}-f,_{y}w,_{y}+f,_{x}w,_{x}\right) }{%
4f\left( z-t\right) ^{2}}  \nonumber \\
R_{xyxy} &=&\frac{f,_{y}^{2}+f,_{x}^{2}-f\left( f,_{xx}+f,_{yy}\right) }{f},
\label{rie}
\end{eqnarray}
and
\[
R_{tt}=\frac{\Delta w}{2f\left( z-t\right) ^{2}},{\ \,\,\,}R_{xx}=-\frac{%
\left( \nabla f\right) ^{2}-f\Delta f}{2f^{2}}{,}
\]
respectively.

Moreover, the harmonicity condition for $\Phi $ implies the last component
vanishes. In fact, when $\mu =$\textit{const}, $\Delta \Phi =0$ implies for $%
f$ that
\begin{equation}
f\Delta f-(\nabla f)^{2}=0.  \label{harmf}
\end{equation}

\begin{itemize}
\item  When $f$ is a constant function, Eqs. (\ref{rie}) reduce to
\begin{equation}
R_{txzx}=\frac{w,_{xx}}{2(z-t)^{2}}{,\,}~~\,R_{txzy}=\frac{w,_{xy}}{%
2(z-t)^{2}}{,\,\,}~~R_{tyzy}=\frac{w,_{yy}}{2(z-t)^{2}}  \label{rief}
\end{equation}
which, $w(x,y)$ being a harmonic function, are all harmonic functions of $%
x,y $. As a consequence, it is straightforward to check that the
generalized Zel'manov criterion, in the form (\ref {gzel}) , is
satisfied \cite{CVV02}.

\item  When $f$ is not a constant function, the generalized Zel'manov
criterion is still satisfied in the form (\ref{gzel}) thanks to a
non trivial combination of the harmonicity condition for $w$ and
Eq. (\ref{harmf}).

\item  In the general case when $f$ and $\mu $ are not constant functions
the Zel'manov criterion is satisfied in the form expressed by
Eq.(\ref{zel}). This may be more conveniently checked in the
adapted coordinates $x^{\mu } $.
\end{itemize}

The Zel'manov criterion, even if it is covariant and allows a clear physical
interpretation in terms of \textit{local} \textit{waves} \textit{of} \textit{%
curvature}, does not determine the propagation direction of the waves, that
is the most important ingredient in the study of their polarization. In the
next sections we will overcome this drawback of the Zel'manov criterion by
using a suitable energy-momentum pseudo-tensor.

Besides the Zel'manov-Zakharov criterion, the Pirani algebraic criterion,
which is based on the Petrov classification, is satisfied. First of all, let
us recall that a vacuum solution of the Einstein equations is a
gravitational wave according to Pirani if its Riemann tensor is of \textit{%
type} \textbf{II}, \textbf{N} or \textbf{III} in the Petrov classification
\cite{Pe69}. Then, in light-cone coordinates ($u=\left( z-t\right) /\sqrt{2}%
,\;v=\left( z+t\right) /\sqrt{2}$), where the metrics given by Eq.(\ref{gw})
read
\begin{equation}
g=2f(dx^{2}+dy^{2})+2dudv+dw~d\ln \left| u\right| ,  \label{presegw}
\end{equation}
the vector fields $\partial _{u}$ and $\partial _{v}$ are both isotropic.
Moreover, it is trivial to show that the only non vanishing components of
the Riemann tensor are
\[
R_{uiuj}=\pm \frac{1}{2u^{2}}\partial _{ij}^{2}w
\]
and this clearly corresponds to a \textit{type}-\textbf{N} Riemann tensor in
the Petrov classification. Furthermore, it follows from the natural
interpretation of the Pirani criterion \cite{Pi57} that the gravitational
wave propagates along the null vector field $\partial _{u}$, or, in other
words, the gravitational wave (\ref{gw}) propagates along the $z-$axis with
velocity $c=1$. Thus, the Pirani criterion, even if with a less clear
physical interpretation, allows an easy and covariant determination of the
propagation direction. It will be an important self-consistency check for
our calculations to discover the same results by means of the
energy-momentum pseudo-tensors.

\section{The sources}

In the theory of gravitational waves a crucial problem is to
characterise realistic sources able to generate waves enough
strong to be detected by the experimental devices.

The simplest source for metrics (\ref{gm}) (with $\sigma =1$) is \textit{dust%
} with density $\rho $ and velocity\textit{\ }$U^{\mu }$\textit{\ }and, then,%
\textit{\ }characterized by an energy-momentum tensor $T_{\mu \nu
}=\rho U_{\mu }U_{\nu }$. When $U^{\mu }$ is a light-like vector
field, this kind
of energy-momentum tensor can describe the energy and momentum of \textit{%
null electromagnetic waves}, \textit{i.e.,} electromagnetic fields
whose scalars, $F^{\mu \nu }F_{\mu \nu }$ and $\epsilon ^{\alpha
\beta \mu \nu }F_{\alpha \beta }F_{\mu \nu }$, both vanish. In
this way it would be possible to describe the gravitational
effects (in particular the emission
of gravitational waves) of a very interesting astrophysical phenomenon, the $%
\gamma -$ray bursts (GRBs): emission of ultra-high energetic $\gamma $-rays (%
$\sim 10^{20}$ $eV$), whose origin is still to be fully
understood.

Moreover, with this source one can also preserve the symmetries
\cite {SVV00,CVV02,CV02b} of the vacuum solution. It can be shown
\cite{CV02b} that, in the adapted coordinates $\left(
x^{1},x^{2},x^{3},x^{4}\right) $, the corresponding solutions of
Einstein equations represent the following gravitational fields
\[
g=2f(dx^{1}dx^{1}+dx^{2}dx^{2})+\mu \lbrack (\widetilde{w}\left(
x^{1},x^{2}\right) -2x^{4})dx^{3}dx^{3}+2dx^{3}dx^{4}],
\]
where $\mu =D\Phi +B$; $D,B\in \mathcal{R}$, $\Phi $ is a non
constant
harmonic function of $x^{1}$ and $x^{2}$, $f=\left( \nabla \Phi \right) ^{2}%
\sqrt{\left| \mu \right| }/\mu $, while $\widetilde{w}\left(
x^{1},x^{2}\right) $ is a solution of the $\mu $-\textit{deformed
Poisson equation}
\[
{\mu }\Delta \widetilde{w}+\nabla {\mu }\cdot \nabla \widetilde{w}=2f%
{\mu }^{2}\rho ,
\]
in which, to save writing, beyond $c=1$, it has been taken $8\pi
G=1$, $G$ being the Newton gravitational constant.

It is worth to remark that spin-1 gravitational waves in vacuum
necessarily are not square integrable. Thus, if they are requested
to be asymptotically flat, then $\delta $-like singularities
appear in the plane $(x,y)$ in a neighborhood of the origin; in
presence of matter sources the singularities are smoothed out.

\section{The energy-momentum pseudo-tensors}

The definition of \textit{momentum} and \textit{energy }associated
with a gravitational field is an intrinsically controversial
problem because these quantities are connected to the space-time
translation invariance, whereas the group of invariance of general
relativity is much bigger. With this cautionary remark in mind,
various definitions are available which attain to different
physical situations. When dealing with the solutions of the
linearised Einstein equations in the vacuum (plane gravitational
waves) a commonly accepted definition is based on the
\textit{canonical} energy-momentum pseudo-tensor (
\cite{Di75,We72}):
\begin{equation}
\tau _{\mu }^{\nu }=\frac{\partial L}{\partial \left( \partial
_{\nu }g_{\alpha \beta }\right) }\partial _{\mu }g_{\alpha \beta
}-g_{\mu }^{\nu }L, \label{dirac}
\end{equation}
where
\begin{equation}
L/\sqrt{\left| g\right| }=g^{\mu \nu }\left[ \Gamma _{\mu \nu }^{\lambda
}\Gamma _{\lambda \sigma }^{\sigma }-\Gamma _{\mu \rho }^{\sigma }\Gamma
_{\nu \sigma }^{\rho }\right]  \label{trs}
\end{equation}
is the \textit{Ricci scalar} deprived of terms containing the second
derivative of the metric.

Then, for the wave solutions of the linearised Einstein equations
the energy density $\tau _{0}^{0}$ is expressed \cite{Di75} as the
sum of squares of derivatives of some metric components which do
represent the physical degrees of freedom of the metric.

Under a transformation preserving the propagation direction and the harmonic
character of the coordinates system, in particular a rotation in the $\left(
x,y\right) $ plane, the physical components of the metric transform like a
spin-$2$ field. It is well known that in general $\tau _{\mu }^{\nu }$ in
Eq. (\ref{dirac}) is not a tensor field but it does transform as a tensor
field under those transformations which preserve the character of the field
of consisting only of waves moving in the $z$ direction, so that the $g_{\mu
\nu }$ remain functions of the single variable $z-t$.

Thus, within the linearised theory, the \textit{canonical}
energy-momentum pseudo-tensor is a good tool to study the physical
properties of the gravitational waves.

\subsection{Comparison with the linearised theory}

The exact gravitational wave
\begin{equation}
g=dx^{2}+dy^{2}+dz^{2}-dt^{2}+d\left( w\right) d\left( \ln \left| z-t\right|
\right) ,  \label{segw}
\end{equation}
given by Eq. (\ref{gw}) for $\mu =1,\,\,f=1/2$ , has the
physically interesting form of a \textit{perturbed Minkowski
metric } with $h=dwd\ln \left| z-t\right|$ .

Moreover, besides being an exact solution of the Einstein equations, it is a
solution of the linearised Einstein equations on a flat background too:
\[
\left\{
\begin{array}{l}
\eta ^{\mu \nu }\partial _{\mu }\partial _{\nu }h=0 \\
\eta ^{\mu \nu }(2h_{\mu \rho ,\nu }-h_{\mu \nu ,\rho })=0
\label{wh}
\end{array}
\right.
\]

Then, to study its energy and polarization, the standard tools of
the linearised theory and in particular the \textit{canonical}
energy-momentum pseudo-tensor, could be used. Nevertheless, with
$h=d\left( w\right) d\left( \ln \left| z-t\right| \right)$  the
$\tau _{0}^{0}$ component of the canonical energy-momentum tensor
vanishes. This is due to the fact that the components of the
tensor $h$ cannot be expressed in the transverse-traceless gauge
since $h$ has only one index in the plane transversal to the
propagation direction.

Even if not explicitly declared, the standard textbooks analysis
of the polarization is performed for the \textit{square
integrable} solutions of the wave-equation. Indeed, they can be
always Fourier developed in terms of plane-wave functions with a
\textit{light-like} vector wave $k_{\mu }$.

The harmonicity condition for the plane wave solutions $h_{\mu \nu
}=e_{\mu \nu }e^{i\rho }+e_{\mu \nu }^{*}e^{-i\rho }$ with $\rho
=k_{\mu }x^{\mu }$ and $k_{\mu }k^{\mu }=0$, reduces to
\begin{equation}
{\frac{1}{2}}k_{\lambda }\eta ^{\mu \nu }e_{\mu \nu }=\eta ^{\mu \nu }k_{\nu
}e_{\mu \lambda }.  \label{harm3}
\end{equation}
It is trivial to see that the symmetry group of this equation,
which encodes the harmonic nature of the coordinate system,
reduces to linear transformations and more precisely to
Poincar\'{e} transformations \cite {We72} (these are nothing but
the usual ''gauge transformations'' of the linearised gravity). It
can be easily shown that, for \textit{square integrable}
perturbations, one can always choose the transverse-traceless
gauge. In other words, it is always possible to eliminate, with a
suitable gauge transformation, the components of the perturbations
with one index in the propagation direction \cite{We72},
\textit{i.e.} \textit{square integrable} perturbations of spin-1
do not exist. For these reasons, the \textit{canonical}
energy-momentum pseudo-tensor, which is gauge invariant in the
sense of the linearised gravity, ''cannot see'' the energy and
momentum of gravitational fields given by Eq.(\ref{segw}) which
have one index in the
propagation direction\footnote{%
Of course, it is possible to find \cite{St90} a coordinate system
in which the perturbation $h$\thinspace\ has non vanishing
components only in the transverse plane. However such a coordinate
system is not harmonic.} (but they are not \textit{square
integrable} so they cannot be gauged away). The fact that
\textit{square integrable} perturbations with one index in the
propagation direction are always \textit{pure} \textit{gauge} is
equivalent to the fact that, for such perturbations, the canonical
energy-momentum pseudo-tensor identically vanishes.

\subsection{Landau--Lifshitz's and Bel's energy-momentum pseudo-tensors}

Besides the canonical energy-momentum pseudo-tensor, a deep
physical significance can be given to the Landau-Lifshitz
energy-momentum pseudo-tensor $\tau ^{\rho \kappa }$ \cite{LL}
defined by
\begin{eqnarray}
\tau ^{\rho \kappa } &=&\frac{1}{16\pi k}\left\{ (2\Gamma _{\lambda \mu
}^{\nu }\Gamma _{\nu \sigma }^{\sigma }-\Gamma _{\lambda \sigma }^{\nu
}\Gamma _{\mu \nu }^{\sigma }-\Gamma _{\lambda \nu }^{\nu }\Gamma _{\mu
\sigma }^{\sigma })(g^{\rho \lambda }g^{\kappa \mu }-g^{\rho \kappa
}g^{\lambda \mu })\right.  \nonumber \\
&+&g^{\rho \lambda }g^{\mu \nu }(\Gamma _{\lambda \sigma }^{\kappa }\Gamma
_{\mu \nu }^{\sigma }+\Gamma _{\mu \nu }^{\kappa }\Gamma _{\lambda \sigma
}^{\sigma }-\Gamma _{\nu \sigma }^{\kappa }\Gamma _{\lambda \mu }^{\sigma
}-\Gamma _{\lambda \mu }^{\kappa }\Gamma _{\nu \sigma }^{\sigma })  \nonumber
\\
&+&g^{\kappa \lambda }g^{\mu \nu }(\Gamma _{\lambda \sigma }^{\rho }\Gamma
_{\mu \nu }^{\sigma }+\Gamma _{\mu \nu }^{\rho }\Gamma _{\lambda \sigma
}^{\sigma }-\Gamma _{\nu \sigma }^{\rho }\Gamma _{\lambda \mu }^{\sigma
}-\Gamma _{\lambda \mu }^{\rho }\Gamma _{\nu \sigma }^{\sigma })  \nonumber
\\
&+&\left. g^{\mu \lambda }g^{\sigma \nu }(\Gamma _{\lambda \nu }^{\rho
}\Gamma _{\mu \sigma }^{\kappa }-\Gamma _{\lambda \mu }^{\rho }\Gamma _{\nu
\sigma }^{\kappa })\right\} .  \label{landau}
\end{eqnarray}

There are strong evidences that, in some cases, it gives the correct
definition of energy \cite{PP79}. In fact, the energy flux radiated at
infinity for an asymptotically flat space-time, evaluated with the
Landau-Lifshitz energy-momentum pseudo-tensor, has been seen to agree with
the Bondi flux \cite{BVM62} that is with the energy flux evaluated in the
exact theory.

It is easy to check that the components $p^{\mu }\equiv \tau _{0}^{\mu }$ of
the $4$-momentum density are
\[
\left\{
\begin{array}{l}
p^{0}=\frac{4}{\left( t-z\right) ^{2}}[C_{1}(w_{,xx})^{2}+C_{2}(w_{,xy})^{2}%
]+\frac{4}{\left( t-z\right) ^{4}}C_{3}\nabla {\ensuremath{\cdot}}[\left|
\nabla w\right| ^{2}\nabla w], \\
p^{1}=p^{2}=0, \quad p^{3}=p^{0},
\end{array}
\right.
\]
where $C_{i}$ are some positive numerical constants, $\nabla =\left(
\partial _{x},\partial _{y}\right) $ and the harmonicity condition for $w$
has been used.

The use of the Bel's superenergy tensor \cite{Be58}
\[
T^{\alpha \beta \lambda \mu }=\frac{1}{2}\left( R^{\alpha \rho \lambda
\sigma }R_{\,\,\,\,\rho \;\;\sigma }^{\beta \;\;\mu }+\left. ^{\ast
}R\right. ^{\alpha \rho \lambda \sigma }\left. ^{\ast }R\right.
_{\,\,\,\,\rho \;\;\sigma }^{\beta \;\;\mu }\right) ,
\]
where the symbol $\ast $ denotes the \textit{volume dual}, leads to the same
result. Indeed, in adapted coordinates the metric has the form
\[
g=dx^{1}dx^{1}+dx^{2}dx^{2}+(w\left( x^{1},x^{2}\right)
-2x^{4})dx^{3}dx^{3}+2dx^{3}dx^{4}
\]
and the only non vanishing independent components of the covariant Riemann
tensor \ $R_{\alpha \beta \gamma \delta }=g_{\alpha \rho }R_{\,\,\,\,\,\beta
\gamma \delta }^{\rho }$ are
\[
R_{1313}=-w,_{11};\,\,\,\,R_{\,1\,323}=-w,_{12};\,\,\,\,\,R_{2323}=-w,_{22}.
\]
It follows that the density energy represented by the Bel's scalar
\[
W=T_{\alpha \beta \lambda \mu }U^{\alpha }U^{\beta }U^{\lambda }U^{\mu },
\]
the $U^{\alpha }$'s denoting the components of a time-like unit vector
field, depends on the squares of $w,_{ij}$. Thus, both the Landau-Lifshitz
pseudo tensor and the Bel superenergy tensor single out the same physical
degrees of freedom. In particular, we can take the components $h_{tx}$ and $%
h_{ty}$ as fundamental degrees of freedom for the gravitational wave (\ref
{segw}).\qquad

Concerning the definition of the polarization, the above form for
$\tau _{0}^{\mu }$ is particularly appealing because, apart from a
physically irrelevant total derivative that does not contribute to
the total energy flux, the component $\tau _{0}^{0}$ representing
the energy density is expressed as the sum of square amplitudes.
The momentum $p^{i}=\tau _{0}^{i}$ is non vanishing only in the
$z-$direction and it is proportional to the energy with
proportionality constant $c=1$; that is these waves move with
light velocity along the $z-$axis. Moreover, this result is
perfectly consistent with the one obtained with the Pirani
criterion.

\section{Spin}

The definition of \textit{spin} or \textit{polarization} for a
theory, such as general relativity, which is non-linear and
possesses a much bigger invariance than just the Poincar\'{e} one,
deserves a careful analysis \cite{Ri98}.

It is well known that the concept of particle together with its degrees of
freedom like the spin may be only introduced for linear theories (for
example for the Yang-Mills theories, which are non linear, it is necessary
to perform a perturbative expansion around the linearised theory). In these
theories, when Poincar\'{e} invariant, the particles are classified in terms
of the eigenvalues of two Casimir operators of the Poincar\'{e} group, $%
P^{2} $ and $W^{2}$ where $P_{\mu }$ are the translation generators and $%
W_{\mu }={\frac{1}{2}}\epsilon _{\mu \nu \rho \sigma }P^{\nu }M^{\rho \sigma
}$ is the \textit{Pauli-Ljubanski polarization vector} with $M^{\mu \nu }$
Lorentz generators.

As it has been shown, the gravitational fields of Eq. (\ref{segw}) represent
gravitational waves moving at the velocity of light, that is, in the would
be quantised theory, particles with zero rest mass. Thus, if a
classification in terms of Poincar\'{e} group invariants could be performed,
these waves would belong to the class of unitary (infinite-dimensional)
representations of the Poincar\'{e} group characterized by $P^{2}=0$, $%
W^{2}=0$. But, in order for such a classification to be meaningful, $P^{2}$
and $W^{2}$ have to be invariants of the theory. This is not the case for
general relativity, unless we restrict to a subset of transformations
selected for example by some physical criterion or by experimental
constraints. For the solutions of the linearised vacuum Einstein equations
the choice of the harmonic gauge does the job \cite{We72}. There, the
residual gauge freedom corresponds to the sole Lorentz transformations.

For these reasons, only gravitational fields represented by Eq. (\ref{segw})
will be considered, which, besides being exact solutions, solve the
linearised vacuum Einstein equations as well. There exist several equivalent
procedures to evaluate their polarization. For instance, one can look at the
$\tau _{0}^{0}$ component of the Landau-Lifshitz pseudo-tensor and see how
the metric components that appear in $\tau _{0}^{0}$ transform under an
infinitesimal rotation $\mathcal{R}$ in the plane $\left( x,y\right) $
transverse\footnote{%
With respect to the Minkowskian background metric, this plane is orthogonal
to the propagation direction. With respect to the full metric this plane is
transversal to the propagation direction and orthogonal only in the limit $%
\left| z-t\right| \mapsto \infty $.} to the propagation direction\footnote{%
It has been said before, that this transformation preserves the harmonicity
condition.}.

The physical components of the metric are $h_{tx}$ and $h_{ty}$ and under
the infinitesimal rotation $\mathcal{R}$ in the plane $\left( x,y\right) $
transform as a vector. Applied to any vector $\left( v_{1},v_{2}\right) $
the infinitesimal rotation $\mathcal{R}$, has the effect
\[
\mathcal{R}v_{1}=v_{2}\,,\,\,\,\,\,\mathcal{R}v_{2}=-v_{1},
\]
from which
\[
\mathcal{R}^{2}v_{i}=-v_{i}\,\,\,\,\ i=1,2,
\]
so that $i\mathcal{R}$ has the eigenvalues $\pm 1$.

Thus, the components of $h_{\mu \nu }$ that contribute to the energy
correspond to spin-$1$ fields\footnote{%
There is nothing to forbid the existence of two spin-1 fields, but
one consequence is that particles with the same orientation repel
and particles with opposite orientation attract.}, provided that
only Lorentz transformations are allowed.

Spin-$0$ and spin-$1$ gravitons have been considered, in a
different context, in \cite{KA02,ADK02,Pa65,SC01}.

\section{Conclusions}

It has been shown that gravitational fields (\ref{presegw})
represent spin-1 gravitational waves and that the reason why it is
commonly believed that spin-1 gravitational waves do not exist is
that, in dealing
with the linearised Einstein theory, all authors implicitly assume a \textit{%
square} \textit{integrable} perturbation. In other words,
\textit{square} \textit{integrable} spin-1 gravitational waves are
always \textit{pure} \textit{gauge}. However, it has been proven
that there exist interesting \textit{non square integrable}
wave-like solutions of linearised Einstein equations that have
spin-1. These solutions are very interesting at least for two
reasons. Firstly, they are asymptotically flat (with at least a
$\delta$-like singularity) in the plane orthogonal to the
propagation direction. Secondly, they are solutions of the exact
equations too, so that the spin-1 cannot be considered as an
''artifact'' of the linearised theory. Realistic sources able to
smooth out the mentioned singularities  have also been found.


\begin{thebibliography}{99}

\bibitem{ADK02} D. V. Ahluwalia, N. Dadhich, M. Kirchbach, \textit{Int. J. Mod. Phys. }\textbf{D} (2002)1621%


\bibitem{KA02} M. Kirchbach, D. V. Ahluwalia, \textit{Phys. Lett.} \textbf{B 529} (2002) 124%

\bibitem{Be58}  L. Bel, \textit{Compt. Rend. Acad. Sci. Colon.} 247 (1958)1094.%

\bibitem{BZ78}  V. A. Belinsky and V. E. Zakharov, \textit{Sov. Phys. Jetp}
\textbf{48}, 6 (1978);\textbf{\ 50}, 1 (1979).

\bibitem{BVM62}  H. Bondi, M.G. J. van der Burg, A. W. K. Metzner, \textit{\
Proc. Roy. Soc.} {A269}, 21 (1962).

\bibitem{CV02b}  F. Canfora and G. Vilasi, \textit{Spin-1 gravitational
waves from cosmic strings and }$\gamma $\textit{-ray bursts}, gr-qc/0301083

\bibitem{CVV02}  F. Canfora, G. Vilasi and P. Vitale, \textit{Phys. Lett.}
\textbf{B 545}(2002)373.

\bibitem{Ch91}  D. Christodoulou, \textit{Phys. Rev. Lett.} \textbf{67},
1486 (1991)

\bibitem{Di75}  P. A. M. Dirac, \textit{General Theory of Relativity} (J.
Wiley \& Sons, N. Y., 1975).

\bibitem{ER37}  A. Einstein and N. Rosen, \textit{J.Franklin Inst.} \textbf{%
223}, 43 (1937).

\bibitem{KKK99}  M. O. Katanaev, T. Kl\"{o}sch, W. Kummer, \textit{Annals
Phys.} 276 (1999)191.

\bibitem{Ko58}  A. S. Kompaneyets, \textit{Sov. Phys. JETP} \textbf{7}, 659
(1958).

\bibitem{KSHM80}  D. Kramer, H. Stephani, E. Herlt, M. MacCallum, \textit{%
Exact solutions of Einstein field equations} Cambridge University Press 1980

\bibitem{LL}  L. D. Landau and E. E. Lifshitz, \textit{Theorie du champ,}
Mir Moscow 1965.

\bibitem{NN01}  H. Nicolai, A. Nagar, \textit{Infinite-dimensional
symmetries in gravity}, in \textit{Gravitational Waves, }IOP (2001) ( I.
Ciufolini et al. eds.)

\bibitem{Pa65}  G. Papini, \textit{Il Nuovo Cimento} \textbf{39}, 716 (1965)

\bibitem{Pe59}  A. Peres, \textit{Phys. Rev. Lett.} \textbf{3}, 571 (1959)]

\bibitem{Pe69}  A. Z. Petrov, \textit{Einstein spaces, (}Pergamon Press,
\textit{\ }New York 1969).

\bibitem{PP79}  S. Persides and D. Papadopoulos, \textit{Gen. Rel. Grav.}
\textbf{\ \ 11}, 233 (1979).

\bibitem{Pi57}  F.\ Pirani, \textit{Phys. Rev.} \textbf{105}, 1089 (1957)

\bibitem{Ri98}  A. Rizzi, \textit{Phys. Rev Lett. \textbf{81}, 6 1150 (1998)}
; \textit{Phys. Rev D}. 63, 104002 (2001).

\bibitem{Ro54}  N. Rosen, \textit{Bull. Res. Coun. Isr.} \textbf{3}, 328
(1954).

\bibitem{SC01}  M. L. Schmid, \textit{Coupling of vector fields at high
energies, }hep-th/9911250 v3 (2002).

\bibitem{St90}  H. Stephani, \textit{General relativity}, Cambridge
University Press 1990

\bibitem{SVV00}  G. Sparano, G. Vilasi and A.M.Vinogradov, \textit{Phys.
Lett.} \textbf{B 513}, 142 (2001); \textit{Differ. Geom. Appl.}
\textbf{16}, 95 (2002);\textbf{\ 17},15 (2002)

\bibitem{Th95}  K. S. Thorne, \textit{Gravitational waves from compact bodies%
} gr-qc/9506084; \textit{Gravitational waves }gr-qc/9506086

\bibitem{Ve93}  E. Verdaguer, \textit{Physics Reports }229, n.1 \& 2,\textit{%
\ 1-80 (1993).}

\bibitem{VV01}  G. Vilasi and P. Vitale, \textit{Class. Quant. Grav.} \textbf{19 }(2002)1-8
(ESI preprint 1110, gr-qc/020218),

\bibitem{Wa84}  R. Wald, \textit{General Relativity} (University of Chicago
Press, Chicago, 1984)

\bibitem{We72}  S. Weinberg, \textit{Gravitation and Cosmology} (J. Wiley \&
Sons, N. Y., 1972).

\bibitem{Za72}  V. D. Zakharov, \textit{Gravitational waves in Einstein's
theory,} (Halsted Press, N.Y.1973)
\end{thebibliography}
\end{document}